\documentclass[oneside,twocolumn,aps,pra,preprintnumbers,superscriptaddress,amsmath,amssymb,10pt]{revtex4-2}

\usepackage{graphicx}
\usepackage{subfigure}
\usepackage{dcolumn}
\usepackage{braket}
\usepackage{array}
\usepackage{booktabs}
\usepackage{bm}
\usepackage[colorlinks=true,
            linkcolor=red,    
            citecolor=red,    
           ]{hyperref} 
\usepackage{amsfonts}
\usepackage{tikz}
\usepackage{xcolor}
\usepackage{float}
\usepackage{quantikz}
\usepackage{csquotes}
\usepackage{soul}

\begin{document}

\title{Conditioning in Generative Quantum Denoising Diffusion Models}

\author{Daniel Quinn}
\email{dquinn53@qub.ac.uk}
\affiliation{Centre for Quantum Materials and Technology, School of Mathematics and Physics, Queen’s University Belfast, Belfast, United Kingdom}

\author{Lorenzo Buffoni}
\affiliation{Department of Physics and Astronomy, University of Florence, Sesto Fiorentino, Italy}

\author{Stefano Gherardini}
\affiliation{Istituto Nazionale di Ottica del Consiglio Nazionale delle Ricerche (CNR-INO), Firenze, Italy}
\affiliation{European Laboratory for Non-linear Spectroscopy, Università di Firenze, Sesto Fiorentino, Italy}

\author{Gabriele De Chiara}
\affiliation{Centre for Quantum Materials and Technology, School of Mathematics and Physics, Queen’s University Belfast, Belfast, United Kingdom}
\affiliation{Física Teòrica: Informació i Fenòmens Quàntics, Departament de Física, Universitat Autònoma de Barcelona, Bellaterra, Spain}

\date{\today}

\begin{abstract}
Quantum denoising diffusion models have recently emerged as a powerful framework for generative quantum machine learning. In this work, we extend these models by introducing a conditioning mechanism that enables the generation of quantum states drawn from multiple target distributions. By sharing parameters across distinct classes of quantum states, our approach avoids the need to train separate models for each distribution. We validate our method through numerical simulations that span single-qubit generation tasks, entangled state preparation, and many-body ground state generation. Across these tasks, conditioning significantly reduced the error of targeted state generation by more than an order of magnitude. Finally, we perform an ablation study to quantify the effect of key hyperparameters on the model performance.
\end{abstract}

\maketitle

\section{Introduction}

Quantum machine learning (QML) investigates how quantum algorithms can be leveraged to learn patterns from data more efficiently than their classical counterparts~\cite{biamonte2017quantum, schuld2021machine, schuld2015introduction}. Within this field, \emph{generative} quantum models aim to learn and sample from quantum or classical data distributions, enabling applications in quantum simulation, quantum chemistry, and quantum state preparation~\cite{tian2022recentadvancesquantumneural}. Existing approaches include models such as quantum circuit Born machines~\cite{Benedetti_2019, Liu_2018}, quantum generative adversarial networks~\cite{Lloyd_2018, Hu_2019, PhysRevA.98.012324, PhysRevApplied.16.024051}, and quantum Boltzmann machines~\cite{Amin_2018, crawford2019reinforcementlearningusingquantum, Zoufal_2021}.

More recently, quantum diffusion models have been proposed as a generalisation of classical diffusion models to the quantum domain, with one of the first formulations introduced in~\cite{parigi2024quantum}. Building on this framework, \emph{quantum denoising diffusion} (QDD) models have subsequently emerged as promising candidates for quantum state generation~\cite{cacioppo2023quantumdiffusionmodels, de2024quantumIma}, inspired by the success of classical denoising diffusion models in image creation~\cite{sohldickstein2015deepunsupervisedlearningusing, ho2020denoisingdiffusionprobabilisticmodels, dhariwal2021diffusionmodelsbeatgans}. Several variants have since been proposed, including hybrid classical–quantum schemes and hardware-based implementations~\cite{chen2024quantumgenerativediffusionmodel, Parigi_2024, parigi2025physicsinspiredgenerativeaimodels, zhang2024generative, huang2025continuousvariablequantumdiffusionmodel}, though many of these approaches face challenges in generalising to highly multimodal quantum data distributions.

In classical generative modelling, \emph{conditioning} is a well-established technique that enables a single model to represent multiple target distributions, improving flexibility and reducing the need for separate models~\cite{bao2022conditional}. By contrast, quantum conditioning remains comparatively underexplored. Prior works have typically encoded class labels in ancilla qubits prepared in computational basis states~\cite{cacioppo2023quantumdiffusionmodels}, an inherently discrete and inflexible strategy. Notably, the QDD variant of~\citet{zhang2024generative}, which serves as the baseline for our study, explicitly identified efficient conditioning as an open challenge.

In this work, we introduce the \emph{Conditioned Quantum Denoising Diffusion} (CQDD) model, which extends the QDD framework by incorporating a conditioning mechanism. We show that our CQDD model can successfully learn diverse quantum data distributions, including quantum many-body phases, entangled states, and non-trivial topological ring structures. We benchmark our CQDD against unconditioned QDD models, finding that conditioning reduces the error in targeted state generation by more than an order of magnitude. Finally, we present an ablation study of key hyperparameters to investigate the trade-offs between model complexity and generative performance.

The paper is organised as follows. In Sec.~\ref{sec:model}, we describe the design of the CQDD model, including details regarding circuit implementations. Following this, in Sec.~\ref{sec:loss}, we outline the loss functions used for training, alongside the broader training methodology. We evaluate the CQDD model on various generation problems in Sec.~\ref{sec:numerical}. Additionally, in Sec.~\ref{sec:benchandgen}, we investigate the predictive limitations of the CQDD model at untrained conditioning angles and benchmark against unconditioned models. Finally, in Sec.~\ref{sec:ablation}, we perform an ablation study of the model.

\section{CQDD Model}
\label{sec:model}

We consider the task of generating new quantum states drawn from an unknown distribution, given access only to finite samples. Extending the standard quantum generative modelling problem, we aim to train a single model capable of generating quantum states from multiple distributions, which is not easily achieved by models without conditioning (see Sec.~\ref{subsec:bench}). Each distribution corresponds to a different \emph{class} or family of quantum states. Unlike traditional approaches, which typically require training a separate model for each target distribution, we seek a unified parameter set $\Theta = \{\boldsymbol{\theta}_1, \ldots, \boldsymbol{\theta}_T\}$ that enables the model to generate samples from different distributions. This structure allows the model to selectively generate quantum states from the desired distribution based on an external conditioning input (e.g., a label).

\begin{figure*}[htbp]
    \centering
    \includegraphics[width=\linewidth]{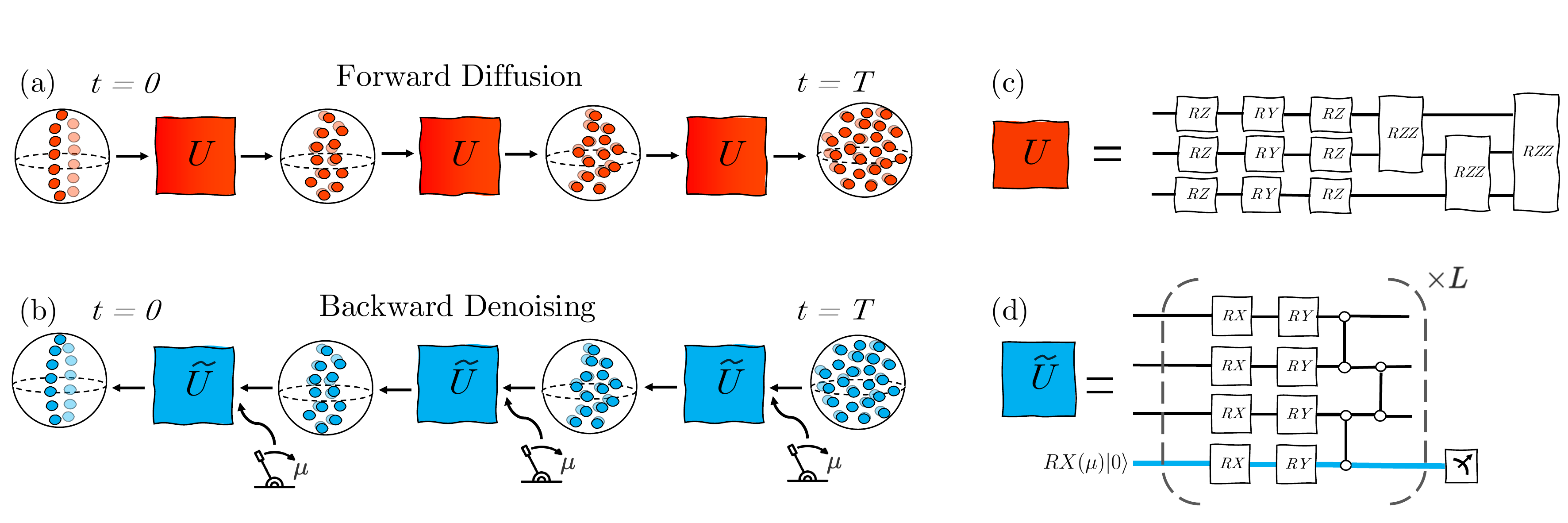}
    \caption{\textbf{Model Schematic.}  
    (a) The forward diffusion process begins with a structured state distribution $S_j(0)$ (left-most Bloch sphere), here in the figure chosen as states distributed along a ring (red dots). Each state is passed through a scrambling unitary $U = U^{(t)}_{i,j}$ (red boxes), and repeated applications of the latter yield a Haar random distribution $S_j(T)$ (right-most Bloch sphere). (b) The reverse denoising process starts from $\tilde{S}_j(T)$ and attempts to reconstruct the original ring distribution. Each state passes through a parametrised unitary $\tilde{U} = \tilde{U}(\boldsymbol{\theta}_t)$ that comprises ancilla qubits, measured and discarded at each backward step. The initial state of the ancilla qubits are conditioned on an angle $\mu$ (represented by the lever). Depending on the number of classes to generate from, we take $\mu$ to be linearly spaced values in the range $0\leq\mu<2\pi$.
    (c) Scrambling circuits use random rotation angles sampled according to Eq.~\eqref{eq:range} for an $n=3$ qubit state. (d) Denoising circuits (example with $n=3$, $n_a=1$) are implemented using an ansatz of $L$ layers of single-qubit rotations and entangling gates. In this circuit, measurements act only on the ancilla qubits initialised as $(RX(\mu)\ket{0})^{\otimes n_a}$, which is illustrated by the thick blue line. The remaining qubits encode the system state $\ket{\tilde{\psi}^{(t)}_{i,j}}$.}
    \label{fig:model}
\end{figure*}

As illustrated in Fig.~\ref{fig:model}, the CQDD model is divided into two parts, similar to the model proposed in~\citep{zhang2024generative}. These two parts represent opposing processes, the \emph{forward diffusion} process and the \emph{backward denoising} process. In the forward process, we begin with a target distribution and gradually add noise, as detailed below, to each state until the distribution becomes Haar random, which we denote as $\mathcal{H}$~\citep{mele2024introduction}. This procedure is applied to each distribution that we aim to model for a total of $T$ steps. In the backward process, a model is trained to reverse this transformation, starting from the random distribution and iteratively refining it into the desired target distribution, conditioned on an external input.

The forward diffusion process is implemented by iteratively applying scrambling circuits~\citep{Belyansky_2020}. At each step $t = 1, \ldots, T$, a unitary $U_{i,j}^{(t)}$ is applied to the $n$-qubit quantum state $\ket{\psi_{i,j}^{(t-1)}}$, where $i$ indexes the $i^{\text{th}}$ sample from the distribution class $j$. Each unitary $U_{i,j}^{(t)}$ is constructed from randomly sampled single- and two-qubit rotations as shown in Fig.~\ref{fig:model}(c). The single qubit rotations are expressed as $RZ(\omega) = e^{-i\omega Z/2}$ which represents a rotation about the $z$-axis by an angle $\omega$, with $Z$ the Pauli-$Z$ operator. A similar definition holds for $RY$ and $RX$ which are related to the Pauli-$Y$ and $X$ operators respectively. The two qubit rotations are given by $RZZ(\omega)=e^{-i\omega Z_{p}Z_q/2}$, applied between every unique pair of qubits $(p,q)$. A subscript on a Pauli operator indicates that it acts only on the specified qubit (e.g., $Z_p$ acts solely on qubit $p$). For all these gates, rotation angles are independently drawn from the uniform distribution
\begin{equation}
\label{eq:range}
\omega\sim\mathcal{U}\left(-\frac{\delta_t \pi}{8},\, \frac{\delta_t \pi}{8} \right),
\end{equation}
where $\delta_t$ is a schedule parameter that controls the noise strength at each diffusion step ($\delta_t$ is discussed in more detail below). Importantly, although the circuit diagram in Fig.~\ref{fig:model}(c) uses the shorthand labels $RZ$, $RY$, and $RZZ$, each occurrence of these rotation gates represents an independent sampling of $\omega$. 

The choice of the interval in Eq.~\eqref{eq:range}, here set to $\pm \delta_t\pi/8$, is arbitrary, as the noise intensity can be adjusted by tuning the schedule $\delta_t$. However, the key requirement is that the final set of states approximates the Haar random distribution $\mathcal{H}$. This iterative process transforms each initial state $\ket{\psi_{i,j}^{(0)}}$ into a progressively more scrambled version $\ket{\psi_{i,j}^{(t)}}$, defined by
\begin{equation*}
\ket{\psi_{i,j}^{(t)}} = \left( \prod_{k=1}^t U_{i,j}^{(k)} \right) \ket{\psi_{i,j}^{(0)}},
\end{equation*}
where the product is ordered such that earlier unitaries are applied first. The resulting set of scrambled states at step $t$ for class $j$ is denoted by $S_j(t) = \left\{ \ket{\psi_{i,j}^{(t)}} \right\}$.

The prefactor $\delta_t$ in Eq.~\eqref{eq:range} depends on the diffusion step $t$ and defines the \emph{noise schedule}, analogous to its role in classical diffusion models~\citep{sohldickstein2015deepunsupervisedlearningusing}. The schedule governs the amount of randomness injected at each step, with $\delta_t$ increasing over time to ensure that the applied rotations progressively explore more of the Hilbert space. This gradual increase in noise drives the set of quantum states towards a Haar random distribution~\citep{chen2023importancenoiseschedulingdiffusion}. If the noise schedule increases too slowly, the final state set may not fully randomise. Conversely, if it increases too rapidly, the states may become effectively random too early in the process, thereby reducing the learnability of the reverse process.

The reverse denoising process is implemented through a sequence of parametrised quantum circuits $\tilde{U}(\boldsymbol{\theta}_k)$ applied at each backward step $k = T, \ldots, 1$~\citep{Benedetti_2019}. The bold notation indicates a vector. Explicitly, $\boldsymbol{\theta}_k \in \mathbb{R}^d$ is a vector of length $d$ consisting of real parameters $\theta_{k,i}$ with $i=1,\dots,d$ for each circuit $\tilde{U}(\boldsymbol{\theta}_k)$. The parameters $\theta_{k,i}$ correspond to the rotation angles of the $RX$ and $RY$ gates, as displayed in Fig.~\ref{fig:model}(d). Beginning with a Haar random initial state $\ket{\tilde{\psi}^{(T)}_{i,j}}$, we compose it with an ancilla register of $n_a$ qubits prepared according to a conditioning angle $\mu$, as detailed below. At each step, the circuit $\tilde{U}(\boldsymbol{\theta}_k)$ acts jointly on the system and ancilla, and is followed by a projective measurement on the ancillary subsystem. This projective measurement realises an effective non-unitary map on the system, approximating the inverse of the forward diffusion step. The ancilla is then discarded, yielding the updated state $\ket{\tilde{\psi}^{(T-1)}_{i,j}}$. Repeating this process across all $T$ reverse steps and for all states yields the reconstructed set at step $t$, denoted by $\tilde{S}_j(t) = \{ \ket{\tilde{\psi}^{(t)}_{i,j}} \}$ for a class $j$.

Each denoising unitary is implemented as an ansatz with $L$ layers of parametrised single-qubit rotations with entangling gates, following the architecture of Ref.~\citep{Kandala_2017}. The entangling gates are controlled-$Z$ operations, $CZ=\mathrm{diag}(1,1,1,-1)$, applied to nearest-neighbour qubits in an alternating pattern. The full circuit is shown in Fig.~\ref{fig:model}(d), where the ancilla qubits are initialised in the state $(RX(\mu)\ket{0})^{\otimes n_a}$. The conditioning angle $\mu$ is defined in the range $0\leq\mu < 2\pi$. For a task consisting of $|\mathcal{C}|$ classes, we take $|\mathcal{C}|$ linearly spaced values of $\mu$ in this range. For example, if $|\mathcal{C}|=2$ then we assign $\mu = 0$ and $\mu=\pi$ to the different classes, or if $|\mathcal{C}|=3$, we assign $\mu = 0, \pi/3$ and $2\pi/3$. The total number of parameters in a single $\tilde{U}(\boldsymbol{\theta}_k)$ is $d=2L(n+n_a)$ and hence in the entire CQDD model we have $2LT(n+n_a)$ trainable parameters. For practical implementations, the circuit depth of the model is $2TL$.

This conditioning mechanism, based on rotations, allows the model to encode more labels than allowed from simply initialising the ancilla in the orthogonal computational basis states (e.g, $\ket{00},\ket{01},\ket{10},\ket{11}$), under the assumption of a fixed number of ancilla qubits $n_a$. Throughout this manuscript, we use $n_a=2$, unless stated otherwise, as we found it to be optimal. However, in practice, $n_a$ is another hyperparameter to tune.

To justify the expressivity and separability of this rotation-based approach, we consider the overlap between the initial conditioning states. Since the ancilla subsystem is prepared via a $RX$ rotation, the state overlap between two distinct conditioning parameters, $\mu$ and $\mu'$, is given by $\cos^2\left((\mu-\mu')/2\right)$; in contrast to orthogonally conditioned states, which strictly enforce zero overlap. Theoretically, the continuous rotation provides the necessary foundation to interpolate between training angles, although it does not inherently guarantee interpolation. Indeed, as our subsequent analysis shows (see Sec.~\ref{subsec:mu}), simply providing a continuous input is, on its own, insufficient to achieve true interpolation.

In the following, we benchmark our rotation-based conditioning strategy against a discrete approach in which each class is assigned a distinct computational basis state of the ancilla register (see Appendix~\ref{app:methods}). Remarkably, we find that rotation-based conditioning yields a lower generative error. This suggests that encoding class information via rotations allows the model to capture class structure more effectively, enhancing the performance of the model at the testing stage with new samples from the initial Haar random distribution $\mathcal{H}$.

\section{Loss Function and Training Methodology}
\label{sec:loss}

To allow for training of the model, one first needs a loss function to minimise, i.e., we require a metric that quantifies the distance between two sets $A$ and $B$ of states. Following the approach in Ref.~\cite{zhang2024generative}, suppose the two sets are finite with $A =\{\ket{\phi_i}\}_{i=1}^N$ and similarly $B=\{\ket{\psi_i}\}_{i=1}^N$ where $N$ is the number of states. Then, we define the pairwise fidelity
\begin{equation}\label{eq:pairwise_fidelity}
    \bar{F}(A,B) = \mathbb{E}_{\ket{\phi} \in A, \, \ket{\psi} \in B}\Big[ F(\psi,\phi) \Big],
\end{equation}
which is based on the simple statewise fidelity $F(\psi,\phi) = \left|\braket{\psi|\phi}\right|^2$. Using the pairwise fidelity (\ref{eq:pairwise_fidelity}), we build the so-called maximum mean discrepancy (MMD) distance~\citep{JMLR:v13:gretton12a}. The MMD distance between the sets can be calculated as
\begin{equation}
    \mathcal{D}_{\rm MMD} (A, B) = \bar{F}(A,A) + \bar{F}(B,B) - 2 \bar{F}(A,B).
\end{equation}
Alternatively, one can use the Wasserstein distance for when it is infeasible to distinguish between two state sets with the MMD distance. Such scenarios show up when one has a distribution that forms a ring, as shown in the Supplementary Material of Ref.~\citep{zhang2024generative}. The Wasserstein distance is computed by casting it into a linear optimisation problem~\citep{annurev-statistics}, which we are going to briefly recall. Suppose, as before, that $A =\{\ket{\phi_i}\}_{i=1}^N$ and similarly $B=\{\ket{\psi_i}\}_{i=1}^N$. Then, we define a $N \times N$ cost matrix $C$, with entries given by the infidelity $C_{i,j} = 1 - \left|\braket{\psi_i|\phi_j}\right|^2$. To compute the Wasserstein distance, we also require a probability distribution governing the sampling of $\ket{\phi_i}\sim A$, and likewise $\ket{\psi_i}\sim B$. Here, as we wish to generate any state from these distributions, we set the probability distributions to be uniform. Hence, we define the Wasserstein distance as the solution to the following linear program:
\begin{equation}
    \mathcal{D}_{\rm WASS}(A, B) = \min_{P} \sum_{i,j}P_{ij} C_{ij},
\end{equation}
subject to $\sum_{j} P_{ij} = \sum_{i} P_{ij} = N^{-1}$ and $P_{ij} \in \mathbb{R}$ with $P_{ij} \geq 0$ for all $i$ and $j$.

Using the MMD or Wasserstein distance, the loss function is defined as the average of the distances computed for each class within the dataset. Mathematically, this is expressed as
\begin{equation}
\label{eq:classloss}
    \mathcal{L}(t) = \frac{1}{|\mathcal{C}|}\sum_{j=1}^{|\mathcal{C}|} \mathcal{D}\left( S_j(t), \tilde{S}_j(t) \right),
\end{equation}
with $|\mathcal{C}|$ denoting the number of classes. This is a relatively simple loss function, as we assume all classes have the same number of states, i.e.~$\left|S_j(t)\right|=|\tilde{S}_j(t)|=N$ for $j=1,...,|\mathcal{C}|$ and all steps $t$. If this were not the case, one could employ a weighted average to counteract the bias of having an unequal number of data. Additionally, we use a rescaled version of the Wasserstein (or MMD) distance, where the distance is divided by $\max_j \mathcal{D}_{\rm WASS}(\mathcal{H}, S_j(0))$. This rescaling accounts for the fact that the distance between the target set $S_j(0)$ and the Haar distribution $\mathcal{H}$ can vary significantly across problem settings, for instance due to differing $n$. As a result, the value $\mathcal{L}(t)$ of the loss function can be interpreted as a relative error, expressed as a percentage with 100\% denoting complete Haar randomness. It is possible for $\mathcal{L}(t)$ to exceed 100\%, in which case the generated set of states should be interpreted as being \textit{worse} than Haar random. For example, if the target is the state $\ket{0}$ and the model generates $\ket{1}$. 

The training methodology of the model follows a \emph{divide-and-conquer} approach. We begin by optimising the parameters of the first backward step, $\boldsymbol{\theta}_T$, to minimise the loss function $\mathcal{L}(T)$. This optimisation is performed using the Adam optimiser with a fixed learning rate~\citep{kingma2017adammethodstochasticoptimization}. The parameters $\boldsymbol{\theta}_T$ are initialised according to a standard Gaussian distribution $\mathcal{N}(0,1)$, with mean $0$ and standard deviation $1$. With these initial values, we generate the corresponding output states from $\tilde{U}(\boldsymbol{\theta}_T)$ and compute the loss $\mathcal{L}(T)$. The Adam optimiser is then applied to update $\boldsymbol{\theta}_T$. After 5000 iterations the optimisation is terminated, and we fix $\boldsymbol{\theta}_T$ to the values that minimise $\mathcal{L}(T)$. 

We then proceed to the denoising circuit $\tilde{U}(\boldsymbol{\theta}_{T-1})$ and compute the loss $\mathcal{L}(T-1)$, again applying the same optimisation procedure. This process is repeated iteratively by progressing backward through time steps until $t=0$, so as to obtain the optimal set of parameters $\Theta = \{\boldsymbol{\theta}_1, \dots, \boldsymbol{\theta}_T\}$. In essence, this training procedure consists of $T$ sequential optimisation problems, each dependent on the parameters learnt in all subsequent steps. More details on optimisation and runtime are discussed in Appendix~\ref{app:grad}.

Both the MMD and Wasserstein distances can be considered \emph{global} loss functions as they are based on state fidelity. Although such loss functions are known to scale unfavourably with the system size due to the exponential decay of fidelity with $n$~\citep{mele2024introduction}, this is not prohibitive in the shallow, small-$n$ regime considered and simulated here. For larger systems, a \emph{local} fidelity-based loss could perform better; however, this lies beyond the scope of the present work~\citep{cerezo2021cost}.

\section{Numerical Simulations}
\label{sec:numerical}

In this section, we evaluate the CQDD model on four diverse datasets to demonstrate its flexibility. In particular, we consider single-qubit states arranged along a ring on the Bloch sphere, clusters of states centred around six poles of the Bloch sphere, Bell states with random relative phases, and ground states of the transverse-longitudinal field Ising model. We also evaluate the performance of the conditioned model against an equivalent unconditioned model, demonstrating the advantage of conditioning for generative tasks. Similar examples for single-class generation (i.e.~no conditioning) are shown in Ref.~\citep{zhang2024generative}. Detailed hyperparameters used in our models for each of these tasks are provided in Appendix~\ref{app:details}.

\subsection{Topological Structure}
\label{subsec:topological}

\begin{figure*}[htbp]
    \centering
    \subfigure{
        \includegraphics[width=0.6\linewidth]{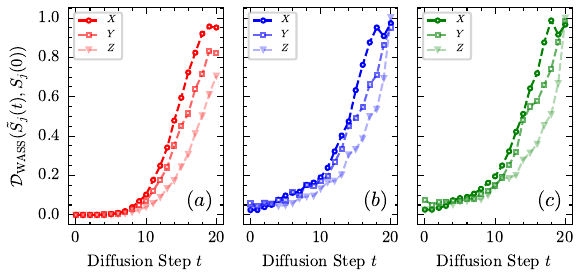}
        \label{fig:rings-norm-distance-intial}
    }
    \subfigure{
        \includegraphics[width=0.28\linewidth]{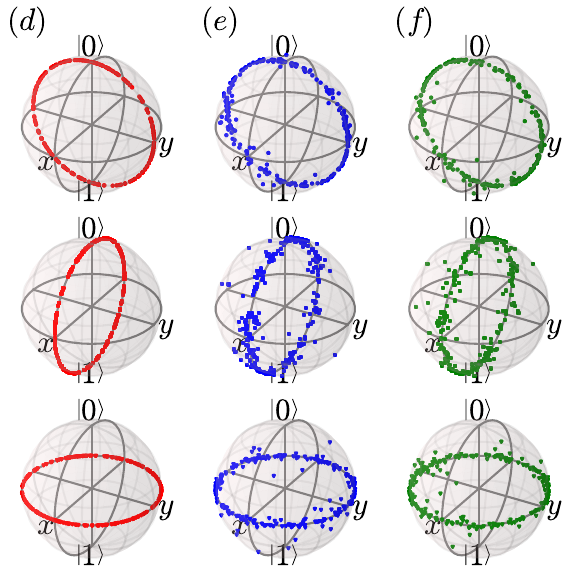}
        \label{fig:rings-bloch}
    }
    \caption{\textbf{Topological Structure Results.} Training and testing performance across all three planar ring classes. The diffusion panel (a) reads left to right, while training (b) and testing (c) panels are read right to left, indicating the backwards processes. The transparency and marker shape denote $S_X$ (dots), $S_Y$ (squares), and $S_Z$ (triangles). In (d)-(f) we depict the Bloch spheres for output states at diffusion step $t=0$. Each row corresponds to $S_X$, $S_Y$, and $S_Z$, and each column to diffusion (red), training (blue), and testing (green). To save on rendering, only the first 250 states are plotted out of $N=1000$.}
    \label{fig:Topological}
\end{figure*}

In the case of a single qubit ($n = 1$), we define three classes of quantum states that the CQDD model aims to regenerate. These classes are defined as follows: 
\begin{align}
    S_X &= \left\{ \cos(\phi) \ket{0} - i \sin(\phi) \ket{1} \quad \text{s.t.} \quad \phi \sim \mathcal{U}(0, 2\pi) \right\}, \label{eq:sx} \\
    S_Y &= \left\{ \cos(\phi) \ket{0} + \sin(\phi) \ket{1} \quad \text{s.t.} \quad  \phi \sim \mathcal{U}(0, 2\pi)\right\}, \label{eq:sy} \\
    S_Z &= \left\{ \left(\ket{0} + e^{-i\phi} \ket{1}\right)/\sqrt{2} \quad \text{s.t.} \quad  \phi \sim \mathcal{U}(0, 2\pi) \right\} \label{eq:sz}.
\end{align}  
Geometrically, $S_X$ corresponds to a ring lying in the $YZ$-plane of the Bloch sphere. Similarly, $S_Y$ and $S_Z$ correspond to rings in the $XZ$- and $XY$-planes, respectively.

Each of these classes undergoes the forward diffusion process using the same noise schedule, where a fixed diffusion factor $\delta_t$ is applied uniformly at every step. Our model conditions the reverse denoising process using a conditioning angle $\mu$ assigned arbitrarily to each ring: $S_X$ is assigned to $\mu = 0$, $S_Y$ to $\mu = 2\pi/3$, and $S_Z$ to $\mu = 4\pi/3$. As a performance metric, we consider the Wasserstein distance  
\begin{equation}
    \mathcal{D}_{\mathrm{WASS}}\big(\tilde{S}_j(t), S_j(0)\big), \quad j \in \{X,Y,Z\}, 
    \label{eq:initial}
\end{equation}  
as shown in Fig.~\ref{fig:Topological}(a)-(c). This distance measures how far the diffused or generated set at step $t$ deviates from its corresponding initial set. In both training and testing, we obtain a comparable final loss of $\mathcal{L}_{\rm WASS}(0) \approx 4.3\%$. To illustrate the fidelity of the learned samples, we visualise the final quantum states on the Bloch sphere in Fig.~\ref{fig:Topological}(d)-(e). Overall, the model performs well in the generation task; the results collectively exhibit the desired ring structure across all three distributions, rather than only one.

\subsection{Polar Points}
\label{subsec:cardinal}

In addition to the planar rings, we also consider a class of $n=1$ qubit states that we refer to as polar points. These states are selected from the six poles of the Bloch sphere along the Cartesian axes, corresponding to the eigenstates of the Pauli operators $X$, $Y$, and $Z$. Explicitly, the states are given by:
\begin{equation}
\mathcal{C} = \Big\{ |0\rangle, |1\rangle, |+\rangle, |-\rangle, |+i\rangle, |-i\rangle \Big\}, \label{eq:cardinal}
\end{equation}
where $|\pm\rangle = \frac{1}{\sqrt{2}}(|0\rangle \pm |1\rangle)$ and $|\pm i\rangle = \frac{1}{\sqrt{2}}(|0\rangle \pm i|1\rangle)$. As we require a set of states, we add Gaussian noise to these eigenstates such that they form clusters around the target directions. That is, for the $\ket{0}$-class, these states are of the form $|\psi\rangle \sim |0\rangle + \epsilon c |1\rangle$, normalised appropriately, where $\text{Re}(c)$ and $\text{Im}(c)$ are independent and follow the Gaussian distribution $\mathcal{N}(0, 1)$. Similar states are sampled for the other classes, such as for the $\ket{+}$-class we have $|\psi\rangle \sim |+\rangle + \epsilon c |-\rangle$, or for the $\ket{-i}$-class we set $|\psi\rangle \sim |-i\rangle + \epsilon c |i\rangle$. The scale factor is set to $\epsilon = 0.08$ and it determines how far the states deviate from the eigenstates. The values of $\mu$ are set to $0, \pi/3, 2\pi/3, \pi, 4\pi/3, 5\pi/3$ and correspond to the eigenstates of $+Z$, $+Y$, $-Z$, $-Y$, $+X$, and $-X$, respectively.

\begin{figure}[htbp]
    \centering
    \includegraphics[width=\linewidth]{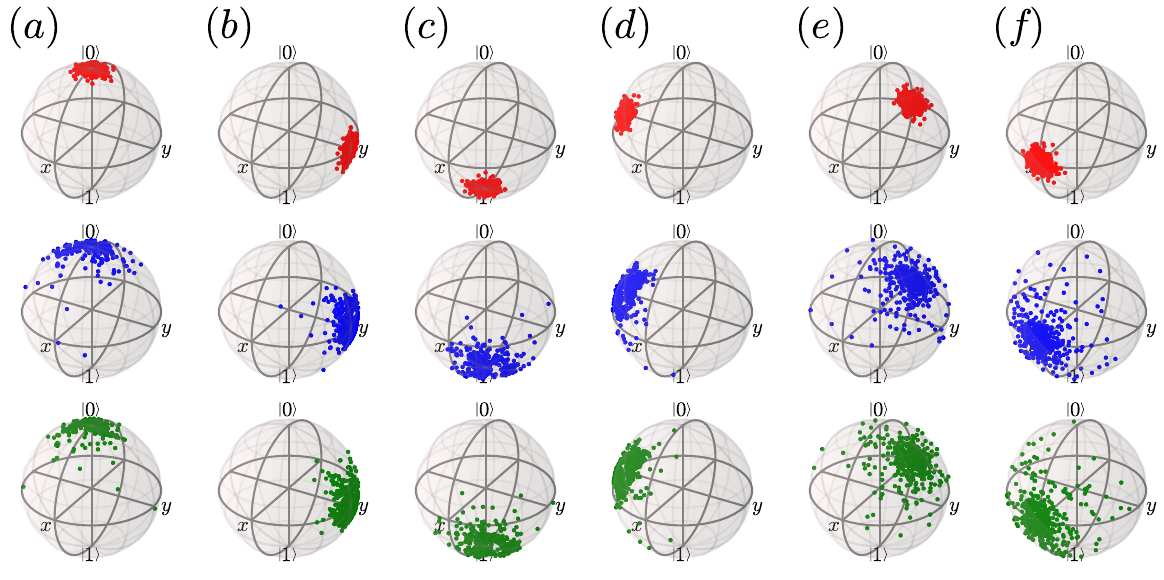}
    \caption{\textbf{Polar Points Results.} Bloch spheres (a)-(f) show the final output states at diffusion step $t = 0$. Each row corresponds to different data sets: diffusion (red), training (blue), and testing (green). Each column represents one of the six cardinal directions (poles) on the Bloch sphere.}
    \label{fig:directions}
\end{figure}

As we now have six total classes, this configuration tests the model’s capacity to correctly generate from a larger number of classes than the discrete basis conditioning method. The limitation arises because with two ancilla qubits ($n_a=2$) there are only $2^{n_a}=4$ computational basis states available, so the method cannot uniquely represent more than four classes. Illustrated in Fig.~\ref{fig:directions} is the final Bloch spheres achieved at $t=0$ for diffusion, training and testing. Note that our model achieves a final test error of $\mathcal{L}_{\rm MMD}(0)\approx1.5\%$.

An alternative figure of merit is the distance of each generated class from its target, analogous to the metric in Eq.~\eqref{eq:initial} but only with the MMD distance. Here, we consider the final step of the denoising process, $t=0$, and evaluate the spread of each Cartesian axis $j \in \{+Z, +Y, -Z, -Y, +X, -X\}$ for the training and testing sets. The rescaled results are reported in Table~\ref{tab:spread} as percentages.
\begin{table}
    \centering
    \setlength{\tabcolsep}{10pt}
    \begin{tabular}{ccc}
        \toprule
        \textbf{Direction} & \textbf{Train} $(\%)$ & \textbf{Test} $(\%)$\\
        \midrule
        $+Z$ & 1.7 & 0.8 \\
        $+Y$ & 1.8 & 2.0 \\ 
        $-Z$ & 1.9 & 2.1 \\ 
        $-Y$ & 1.4 & 1.5 \\
        $+X$ & \textbf{3.7} & \textbf{4.7} \\
        $-X$ & \textbf{5.2} & \textbf{6.4} \\
        \bottomrule
    \end{tabular}
    \caption{\textbf{Distance of Polar Points.} We display values of $\mathcal{D}_{\mathrm{MMD}}\big(\tilde{S}_j(0), S_j(0)\big)/\mathcal{D}_{\mathrm{MMD}}\big(\mathcal{H}, S_j(0)\big)$ for all the directions in percentages. Lower values indicate better agreement between the target distribution class and the training and testing class. Bold values in the train and test columns denote the largest and second-largest values.}
    \label{tab:spread}
\end{table}

From Table~\ref{tab:spread}, the model achieves the smallest spread for the $+Z$ and $-Y$ classes, while the $\pm X$ classes exhibit larger distances, indicating a broader distribution of generated states. This trend is also apparent in Fig.~\ref{fig:directions}, where the training and testing Bloch spheres $(a)$ and $(d)$ are more tightly clustered than $(e)$ and $(f)$. We conjecture that the bias towards $+Z$ and $-Y$ is due to the corresponding conditioning angles being $\mu = 0$ and $\mu = \pi$, which appear to be more favourable for the model’s training dynamics.
 
\subsection{Entanglement Structure}
\label{subsec:entnalge}

Let us now consider generating states with more qubits ($n=2$), where a natural choice is given by the Bell states~\citep{nielsen2010quantum}. To construct a class of entangled states, we introduce a relative phase into the standard Bell states. Specifically, we define two classes:
\begin{align}
    S_\Phi &= \left\{ \frac{\ket{00}+e^{i\phi}\ket{11}}{\sqrt{2}} \quad \text{s.t.} \quad\phi \sim \mathcal{U}(0, 2\pi) \right\}, \label{eq:phi}\\
    S_\Psi &= \left\{ \frac{\ket{01}+e^{i\phi}\ket{10}}{\sqrt{2}} \quad \text{s.t.} \quad\phi \sim \mathcal{U}(0, 2\pi) \right\} \label{eq:psi}.
\end{align}
Each class is identified by a conditioning angle $\mu$, where $\mu = 0$ corresponds to $S_\Phi$ and $\mu = \pi$ to $S_\Psi$. Importantly, these two classes lie in orthogonal subspaces of the Hilbert space, and, unlike the single-qubit case, this two-qubit task requires learning entanglement and correctly switching between distinct subspaces. After training our model, we obtain a final test error of  $\mathcal{L}_{\rm WASS}(0)\approx2\%$. 

As an alternative metric, the degree of entanglement within these sets of states can be quantified via the average Meyer-Wallach measure $\langle Q \rangle$. For a pure normalised state, this measure is defined as
\begin{equation}
    Q = \frac{2}{n} \sum_{i=1}^n \left(1 - \mathrm{Tr}[\rho_{i}^2]\right),
\end{equation}
where \(\rho_{i}\) is the reduced density matrix for qubit \(i\), obtained by tracing out all other qubits~\citep{Meyer_2002,PhysRevA.79.052307}. Our model achieves a value of $\langle Q \rangle=0.97$, compared with the ideal value of 1 for both distributions, since a relative phase does not affect the entanglement measure.

We can also consider a metric that quantifies the \emph{percentage} of a generated state that lies within a specific target subspace. For a normalised two-qubit pure state  
$$
\ket{\psi} = c_0\ket{00} + c_1\ket{01} + c_2\ket{10} + c_3\ket{11},
$$
with $c_i\in \mathbb{C}$, we define the percentage overlap with the $\{\ket{00},\ket{11}\}$ subspace as $|c_0|^2 + |c_3|^2,$ and similarly, the percentage overlap with the $\{\ket{01},\ket{10}\}$ subspace as $|c_1|^2 + |c_2|^2.$ In both cases, we find that any given generated state has an overlap of $99.9\%$ with its respective subspace.

In addition, we investigate the task of generating entangled $n = 3$ qubit states belonging either to the GHZ- or to the W-class~\citep{D_r_2000}. Specifically, we define two target distributions to be a GHZ-class consisting of states of the form $\ket{\mathrm{GHZ}} = (\ket{000} + e^{i\phi}\ket{111})/\sqrt{2}$, and a W-class that includes states of the form $\ket{\mathrm{W}} = (\ket{001} + \ket{010} + e^{i\phi}\ket{100})/\sqrt{3}$, with $\phi$ representing an arbitrary phase. Using a subspace-fidelity metric similar to that introduced earlier, we find that the model successfully generates both GHZ and W states, with approximately 90\% of samples lying within their respective subspaces. However, in contrast to previous experiments, the generated distributions in the Hilbert space do not form well-structured rings. This is reflected in a relatively high final Wasserstein loss of $\mathcal{L}_{\mathrm{WASS}}(0) = 15\%$. We hypothesise that this larger error may stem from the difference in rank between the two target classes as GHZ states span a rank-2 subspace, whereas W states span a rank-3 subspace. Further visualisations and analysis are provided in Appendices~\ref{app:ent} and~\ref{app:rank}.

\subsection{Many-body Phases}
\label{subsec:manybody}

As a physically relevant example, we consider the transverse-longitudinal field Ising model (TLFIM), whose ground states have been studied extensively~\citep{phase-trans}. The model is defined by the Hamiltonian
\begin{equation}
H = - \sum_{i=1}^{n} Z_{i}Z_{i+1} - g \sum_{i=1}^{n} X_i - h \sum_{i=1}^{n} Z_i,
\label{eq:tlfim}
\end{equation}
where periodic boundary conditions are imposed by identifying $Z_{n+1} \equiv Z_{1}$. The transverse magnetic field strength is denoted by $g$, and $h$ represents the longitudinal field. In the absence of the longitudinal field ($h=0$), the system undergoes a quantum phase transition at $g=1$. For $g<1$ the system is in a ferromagnetic phase and when $g>1$ the system enters a paramagnetic phase.

For the state generation task, we focus on $n=4$ qubit ground states drawn from two distinct quantum phases. The transverse field strength is sampled as $g \sim \mathcal{N}(0.5, 0.1)$, a Gaussian distribution with mean $0.5$ and standard deviation $0.1$. Two classes are constructed, with one setting taking the longitudinal field $h=0.25$ and the another setting $h=-0.25$, denoted by $S_+$ and $S_-$ respectively. These classes correspond to states with an incomplete spin alignment in the positive or negative $z$-direction. Furthermore, we set the conditioning angle to $\mu = 0$ for $S_+$ and $\mu = \pi$ for $S_-$. We note that decreasing $|h|$ increases the fidelity between states in $S_+$ and $S_-$, effectively reducing their separation in the Hilbert space. This may render the generation task easier, as the model no longer needs to reproduce states lying at nearly opposite regions in the Hilbert space.

\begin{figure}[htbp]
    \centering
    \includegraphics[width=\linewidth]{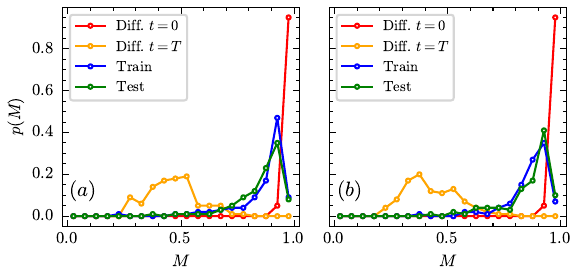}
    \caption{\textbf{Many-body Phase Results.} Probability distributions of magnetisation for generated data from the training set (blue) and testing set (green), compared to the true data distribution (red) and the scrambled distribution at $t=T$ (orange). Panel (a) corresponds to $h = 0.25$, and panel (b) to $h = -0.25$. The shorthand Diff. $t=k$ stands for the set of data at the diffusion step $k$.}
    \label{fig:Manybody}
\end{figure}

To evaluate the fidelity of the generated states, we use the $z$-magnetisation, defined as $M = \sum_{i=1}^{n}Z_i$. Portrayed in Fig.~\ref{fig:Manybody} is the probability distribution $p(M)$ for the magnetisation in both generated and target datasets. A close alignment between the model-generated distributions and the true data is observed, while the diffused states at the end of the forward process are more dispersed, approximating a Gaussian distribution. As another metric, we report the mean magnetisation: the true data yields $\langle M_{\rm targ}\rangle = 0.97$ for both classes, the fully diffused set, i.e. random states, yields $\langle M_{\rm rand}\rangle = 0.43$, and the model outputs achieve $\langle M_{\rm model}\rangle = 0.86$ for both considered classes in training and testing. Additionally, the final test error of the model is $\mathcal{L}_{\rm MMD}(0) \approx 3.7\%$.

\section{Benchmarking and Generalisation}
\label{sec:benchandgen}
In this section, we benchmark our CQDD against its unconditioned counterpart and investigate the models performance at intermediate untrained conditioning angles. Furthermore, Appendix~\ref{app:direct} extends this analysis by comparing the CQDD model with a similarly conditioned quantum direct transport model.

\subsection{Benchmarking of Conditioned and Unconditioned Models}
\label{subsec:bench}

We now highlight a key result of our study, namely the benchmarking of conditioned versus unconditioned QDD models. Benchmarking here serves a dual role, not only as a quantitative performance test, but also as a clear demonstration of the advantage of conditioning in generative quantum modelling.

First, consider a simple $n=1$ qubit task involving the ring structures from Section~\ref{subsec:topological}, where the target distribution is $S = S_X \cup S_Y$ as defined in Eqs.~\eqref{eq:sx} and~\eqref{eq:sy}. Using hyperparameters $T=20$, $n_a=2$, $L=12$, and a total of $N=1000$ unlabelled training states, the unconditioned model reaches a final rescaled test loss of $67.1\%$, indicating poor reconstruction. In contrast, a CQDD model trained with $N=500$ samples per class (also totalling {$1000$}) attains a loss of $4.39\%$, which is more than a $15\times$ improvement. Here, $\mu=0$ generates the $S_X$ class and $\mu=\pi$ generates the $S_Y$ class. The clear difference between loss values illustrates the central advantage of conditioning. Specifically, it enables sampling from either component distribution, rather than a blurred approximation to their union.

We next address a more demanding $n=2$ qubit task of generating maximally entangled states from the union $S = S_{\Phi} \cup S_{\Psi}$, defined in Eqs.~\eqref{eq:phi} and~\eqref{eq:psi}. We take $T=20$, $n_a=2$, $L=12$, and a total of $N=500$ unlabelled training states. The unconditioned model attains a final test loss of $24.1\%$, whereas our CQDD model reduces this to $2.02\%$, which is more than an $11\times$ improvement. Again, $\mu=0$ targets $S_{\Phi}$ and $\mu=\pi$ targets $S_{\Psi}$. These results demonstrate that conditioning systematically improves generative performance.

\begin{figure}[htbp]
\centering
\includegraphics[width=\linewidth]{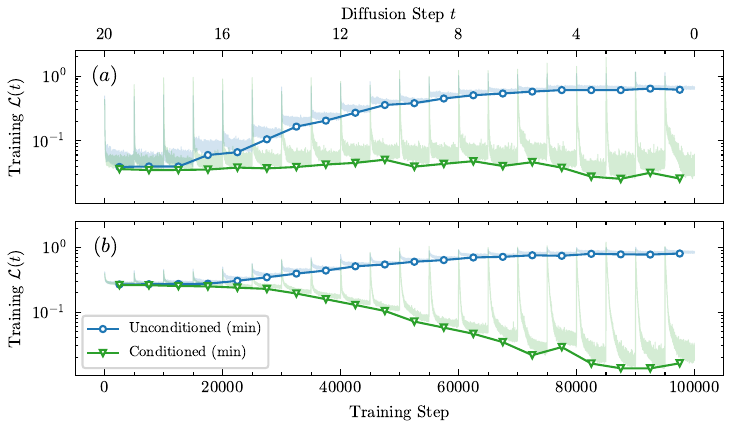}
\caption{\textbf{Benchmarking.} Rescaled training loss for the (a) ring and (b) entanglement generation task. We compare the performance of the unconditioned model (blue) and the conditioned (green) model across training steps. Markers show the minimum loss at each diffusion step $t$, while the faint lines show the raw loss values during the training. Periodic spikes are caused by random initialisation at the start of every step.}
\label{fig:benchmarking}
\end{figure}

In Fig.~\ref{fig:benchmarking}, we show the training loss of both conditioned and unconditioned models for the (a) ring task and (b) entanglement task as described above. The loss for the unconditioned model is given as $2\mathcal{L}(t) = \mathcal{D}(\tilde{S}(t)[:N/2], S(t)[:N/2]) + \mathcal{D}(\tilde{S}(t)[N/2:], S(t)[N/2:])$. Here, we use Python-style notation in which $A(t)[:x]$ denotes the first $x$ elements of the set $A(t)$, while $A(t)[x:]$ denotes the elements from index $x$ to the end. This modified loss function reflects the imposed ordering of the data. In the conditioned model, the parameter $\mu$ explicitly controls which state is generated. Therefore, to ensure a fair comparison, we introduce an analogous partitioning for the unconditioned model by separating the dataset into two subsets. For the conditioned case we have $2\mathcal{L}(t)=\mathcal{D}(\tilde{S}_i(t),S_i(t))+\mathcal{D}(\tilde{S}_j(t),S_j(t))$ where $i=X,\Phi$ and $j=Y,\Psi$ for the respective tasks. Markers in Fig.~\ref{fig:benchmarking} denote the minimum value of the loss for a particular backwards step $t$, which is the value we use to find $\boldsymbol{\theta}_t$. In the faint colours we plot the raw loss value, where the periodic spikes are caused by the randomised initialisation at each backward step.

The conditioned model exhibits a significant reduction in final loss. During approximately the first three steps of the training process, the unconditioned loss matches the conditioned loss in both tasks. However, the training trajectories diverge thereafter: the conditioned loss decreases rapidly, while the unconditioned loss increases, suggesting that optimisation for the unconditioned model fails to converge.

\subsection{Intermediate Untrained Conditioning Angles}
\label{subsec:mu}

It is important to distinguish between the continuous mathematical nature of the conditioning parameter and the model’s actual capacity to generalise across the parameter space. While the parameter $\mu$ is inherently continuous (being an angle $\mu \in [0, 2\pi) \subset \mathbb{R}$), the model is trained only on a discrete subset of these rotation angles. Thus, the model operates as a discrete conditional generator restricted to the trained labels, rather than a true continuous interpolator.

To study this behaviour, we sweep $\mu \in [0, 2\pi]$ in uniform steps and compute the rescaled Wasserstein distance between the generated distribution $G(\mu)$ and its corresponding target set. Two tasks are considered, (i) generation of the state classes $S_X$ and $S_Y$ defined in Eqs.~\eqref{eq:sx} and \eqref{eq:sy}, and (ii) generation of the entangled state classes $S_\Phi$ and $S_\Psi$ defined in Eqs.~\eqref{eq:phi} and \eqref{eq:psi}. In both tasks, $\mu = 0$ and $\mu = \pi$ correspond to the two target classes and both models use $T=20$, $L=12$, $n_a=2$, and $N=125$.

\begin{figure}[htbp]
\centering
\includegraphics[width=\linewidth]{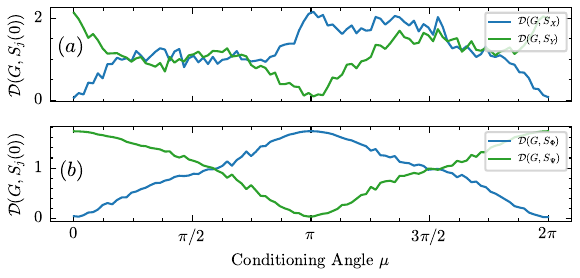}
\caption{\textbf{Generation Distance.} Rescaled Wasserstein distance $\mathcal{D}_{\rm WASS}(G(\mu), S_j(0))$ and the trained target distributions. (a) Distances to $S_X$ (blue) and $S_Y$ (green), where $S_X$ corresponds to $\mu = 0$ and $S_Y$ to $\mu = \pi$. (b) Distances to $S_\Phi$ (blue) and $S_\Psi$ (green) under analogous conditioning.}
\label{fig:mu}
\end{figure} 

\begin{figure}[htbp]
\centering
\includegraphics[width=\linewidth]{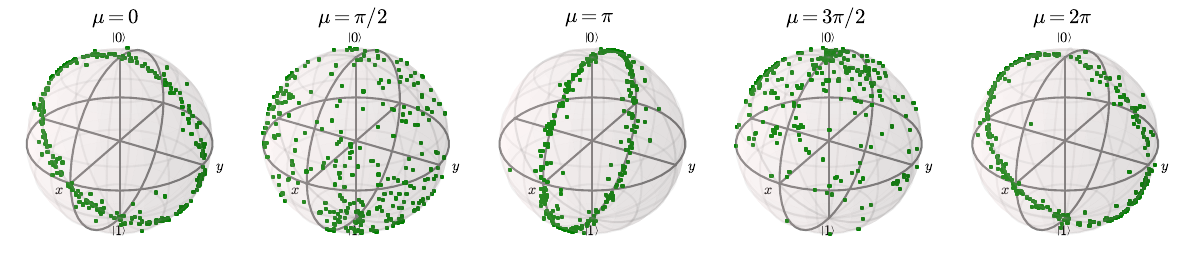}
\caption{\textbf{Intermediate Conditioning.} Bloch-sphere visualisation of $G(\mu)$ for selected values of $\mu$. Each Bloch sphere corresponds to a different conditioning angle. At untrained intermediate angles $\mu = \pi/2$ and $3\pi/2$, the generated states lack the structured rings of the target sets.}
\label{fig:planar-mu}
\end{figure}

For the rings task of Fig.~\ref{fig:mu}(a), the distance to $S_X$ increases with $\mu$ up to a maximum at $\mu = \pi$, while the distance to $S_Y$ follows a complementary trend, such that $\mathcal{D}(G,S_X) + \mathcal{D}(G,S_Y)$ remains approximately constant. This indicates that the model maintains a consistent trade-off between the two classes, producing a distribution close to the intended target as long as $\mu$ is near the desired angle. However, Bloch sphere visualisations, as given in Fig.~\ref{fig:planar-mu}, reveal that at untrained intermediate angles ($\mu=\pi/2,3\pi/2$) the generated states exhibit no resemblance to either target distribution, forming clusters of random points rather than well-defined rings in different planes. Thus, one can regard our model as not truly generalising to all conditioning angles and hence enforcing this remains an interesting open challenge.

A smoother interpolation is observed for the entanglement task of Fig.~\ref{fig:mu}(b), with the crossover of the curves at $\mu \approx \pi/2$ and $\mu \approx 3\pi/2$. At these points, the generated states are nearly equidistant from both target classes, and occupy both subspaces $\{\ket{00},\ket{11}\}$ and $\{\ket{01},\ket{10}\}$ with a 60/40 split favouring $\{\ket{00},\ket{11}\}$. These intermediate states, absent from the training set, suggest that the model can interpolate between entangled targets, albeit without preserving the fine structure of either distribution.

\section{Ablation Study}
\label{sec:ablation}

In this section, we analyse how the model error scales with changes to various hyperparameters. We divide this analysis into two categories: (i) model-dependent scaling, where we vary hyperparameters such as the number of diffusion steps $T$, circuit depth $L$, and number of ancilla qubits $n_a$; and (ii) problem-dependent scaling, where we vary the complexity of the generative task.

\subsection{Model Dependent Scaling}
\label{subsec:model-scale}

\begin{figure}[htbp]
    \centering
    \includegraphics[width=\linewidth]{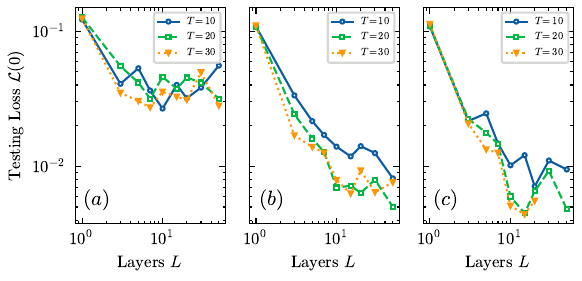}
    \caption{\textbf{Model Scaling.} Testing loss as a function of the number $L$ of layers. We evaluate the impact of the number of diffusion steps $T$ by plotting the results for $T=10$ (blue circles), $T=20$ (green squares), and $T=30$ (yellow triangles). Panel (a) corresponds to $N=100$ data points, (b) to $N=500$, and (c) to $N=1000$.}
    \label{fig:testing-layer}
\end{figure}

We analyse the scaling behaviour of our model, focusing on how its performance varies with the number of data points and the number of layers in the variational circuit. As a benchmark task, we consider the three-class planar ring dataset, introduced in Sec.~\ref{subsec:topological}. The model’s testing loss $\mathcal{L}(0)$ is illustrated in Fig.~\ref{fig:testing-layer} as a function of the circuit depth and is evaluated across three different dataset sizes. We further examine how performance varies with the number of diffusion steps.

The results demonstrate that increasing the number of circuit layers $L$ generally improves the test performance, which is consistent with both classical and quantum learning theory~\citep{sim2019expressibility}. However, beyond a certain depth, the test loss saturates or begins to increase. This is an indication of overfitting, meaning that the model fits the training data too closely but fails to generalise~\citep{bilmes2020underfitting}. This effect is more pronounced for small datasets ($N=100$), where overfitting occurs earlier. For larger datasets ($N=500,1000$), deeper circuits remain beneficial for longer. Similarly, the number of diffusion steps $T$ influences the performance of the models. Larger $T$ enables finer denoising and improves performance, but also introduces more parameters and thus the possibility for overfitting. 

Empirically, we observe that the testing loss exhibits approximate power-law scaling with respect to all three model parameters. In particular, we find behaviour consistent with $\mathcal{L} \propto L^{-3/4}$, $\mathcal{L} \propto N^{-3/4}$, and $\mathcal{L} \propto T^{-1/2}$ over the range of values explored. This indicates diminishing returns as model capacity and dataset size increase, reflecting a trade-off between expressivity and generalisation. We emphasise that these scaling laws are empirical observations based on finite-size simulations, and a more rigorous theoretical understanding remains an open problem.

\begin{figure}[htbp]
    \centering
    \includegraphics[width=\linewidth]{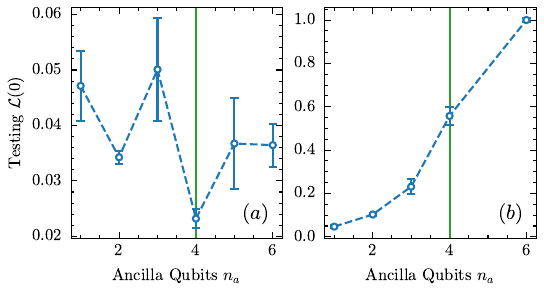}
    \caption{\textbf{Ancilla Scaling.} Final testing loss $\mathcal{L}(0)$ as a function of the number $n_a$ of ancilla qubits for (a) the unconstrained resource regime and (b) the fixed resource regime with constant $L n_a$. In both plots, the green solid lines indicate $n=n_a$.}
    \label{fig:ancilla}
\end{figure}

Another important scaling behaviour worth investigating is the relationship between the model performance and the number of ancilla qubits. To explore this, we consider the learning task of the Ising model introduced in Sec.~\ref{subsec:manybody}. Illustrated in Fig.~\ref{fig:ancilla} is the final scaled testing error as a function of the number of ancilla qubits under two distinct resource regimes. In the \emph{unconstrained} regime, we fix the circuit depth to $L=12$ and vary $n_a$, effectively increasing the total number of trainable parameters. In contrast, in the \emph{constrained} regime, we fix the total number of parameters by keeping the product $L n_a = 12$ constant, adjusting the depth inversely with the ancilla count. The model was evaluated on $N=250$ test states, partitioned into five subsets of 50 samples each. For each subset, we compute the average test loss across both classes. The reported loss is the mean over the partitions, and the error bars indicate the standard deviation of the subsets.

As shown in Fig.~\ref{fig:ancilla}, the two regimes have a clear distinction. In the unconstrained case, increasing $n_a$ initially reduces the testing error, due to an increase in the parameter count. However, error eventually fluctuates, reaching a minimum at $n_a = n = 4$ (marked by the green solid line in both panels of the figure). Interestingly, when $n_a\leq n$, odd ancilla counts ($n_a=1,3$) exhibit higher losses and wider error bars, whereas even ancilla counts ($n_a=2,4$) are lower and more stable. We conjecture that this is due to the entanglement layers in the backwards circuit, as this is where the interplay between the ancilla and system comes into effect. However, a more thorough analysis is needed to determine why this insight breaks down for $n_a>n$, but this goes beyond the scope of this work.

In contrast, the constrained regime shows a clear monotonic increase in the error as $n_a$ increases and the circuit depth decreases. The worst performance occurs at $n_a=6$ and $L=2$, highlighting that for a fixed parameter budget, deeper circuits (larger $L$, smaller $n_a$) are empirically more effective than wider, shallower ones. This aligns with classical deep learning findings, where deep models perform better than wide models~\citep{pleiss2021limitationslargewidthneural, liang2017deepneuralnetworksfunction}.

\subsection{Problem Dependent Scaling}
\label{subsec:problem-scale}

In this section, we investigate how the performance of the CQDD model scales with the problem complexity, specifically in terms of the number of target classes $|\mathcal{C}|$ to be generated. To this end, we consider two representative examples: (i) distributions of single-qubit states that form rings, which are parallel to the equator of the Bloch sphere, and (ii) various 4-qubit GHZ entangled states with varying relative phases. These tasks exemplify low- and high-complexity generative challenges, respectively.

\begin{figure}[htbp]
\centering
\includegraphics[width=\linewidth]{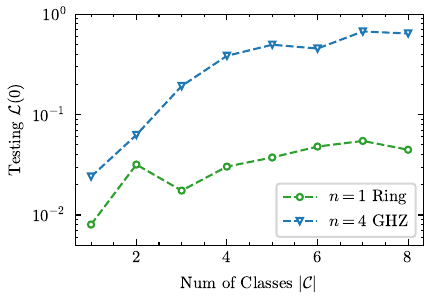}
\caption{\textbf{Class Scaling.} Test loss $\mathcal{L}(0)$ versus the number of classes $|\mathcal{C}|$ for the GHZ (blue circles) and rings (green triangles) tasks.  For both tasks, we have $T=20$ and $n_a=2$. However, for the GHZ task we have $L=15$ and $N=100$, whilst for the rings task $L=12$ and $N=250$.}
\label{fig:class-laws}
\end{figure}

The task \textit{rings}, illustrated in Fig.~\ref{fig:class-laws}, consists of generating single-qubit states drawn from the set
\begin{equation}
\label{eq:alpha}
S(\alpha) = \left\{ \cos\left(\frac{\alpha}{2}\right) \ket{0} + e^{i \phi} \sin\left(\frac{\alpha}{2}\right)\ket{1} \ \text{s.t.}\ \phi \sim \mathcal{U}(0, 2\pi) \right\}
\end{equation}
where each class is associated with a fixed angle $\alpha$. For a problem with $|\mathcal{C}|$ classes, the class-dependent angles are defined as
\begin{equation}
\alpha = \frac{j\pi}{|\mathcal{C}|+1}, \qquad j=1,\dots,|\mathcal{C}|.
\end{equation}
In practice, $S(\alpha)$ contains quantum states whose Bloch vector belongs to a ring parallel to the equator at colatitude $\alpha$.

The task called \emph{GHZ} in Fig.~\ref{fig:class-laws} describes a task where one wishes to generate states from a family of 4-qubit entangled GHZ-type states with varying relative phases. Each class corresponds to 
\begin{equation}
    S (\boldsymbol{x}) = \left\{ \frac{1}{\sqrt{2}} \left(\ket{\boldsymbol{x}} + e^{i \phi} \ket{\bar{\boldsymbol{x}}} \right) \quad \text{s.t.} \quad \phi \sim \mathcal{U}(0, 2\pi)\right\},
\end{equation}
where $\boldsymbol{x}$ is a 4-bit binary string, and $\bar{\boldsymbol{x}}$ denotes its bitwise complement. We take $\boldsymbol{x} \in \{0000,0001,0010,0100,1000,0011,1001,0101\}$ to avoid double-counting equivalent pairs. These GHZ-type states span a structured subspace of the 4-qubit Hilbert space and present a more challenging generative task due to their entanglement and symmetry properties.

Across both tasks, we observe a consistent trend: as the number of target classes $|\mathcal{C}|$ increases, the testing loss $\mathcal{L}(0)$ also increases. However, the rate and scale of degradation differ significantly between the two cases. As shown in Fig.~\ref{fig:class-laws}, the error for the 4-qubit GHZ task is an order of magnitude larger than that for the single-qubit equator task when $|\mathcal{C}| \geq 3$. This disparity highlights how the structure of the target states impacts model performance under fixed computational resources.

\section{Conclusion and Outlook}
\label{sec:conc}

The CQDD model introduced in this work represents a significant advancement toward more powerful and expressive quantum machine learning. By employing a shared set of parameters to generate distinct classes of quantum states, our model eliminates the need to train separate networks for each class. The conditioning mechanism introduces only a minor overhead. Namely, one additional gate on each ancilla qubit, which results in a negligible increase in the circuit depth when compared to the model proposed by~\citet{zhang2024generative}.

We validated the model across four diverse datasets: (i) a set of single-qubit states arranged on a ring on the Bloch sphere, (ii) clusters of states around poles of the Bloch sphere, (iii) two-qubit Bell states with a random relative phase, and (iv) many-body ground states of the transverse-longitudinal field Ising model. These tasks span both abstract and physically motivated distributions, demonstrating the generalisation capabilities and flexibility of the CQDD framework. Benchmarking against an unconditioned model highlighted the benefit of conditioning, yielding an approximate order of magnitude improvement in performance. Furthermore, we conducted an ablation study of key hyperparameters such as the number of denoising steps $T$, circuit depth $L$, and the number of ancilla qubits $n_a$. Finally, we discuss problem-specific factors such as the number of classes to be generated.

Despite the strong performance of the CQDD model, several important questions remain open. During training, conditioning is introduced via a finite set of rotation angles $\mu$ with which the ancilla qubits are initially prepared. When the trained model is then evaluated at untrained intermediate values of $\mu$, the generated states fail to accurately reflect the intended target distribution. One possible explanation is that because the conditioning signal is injected solely through the ancilla subsystem, the signal only indirectly influences the system. A promising direction for future work is to incorporate the conditioning more directly into the system, for example, by embedding $\mu$ throughout the circuit via repeated parametrised operations. This approach is analogous to data re-uploading strategies in quantum neural networks, where inputs are injected at multiple layers to enhance expressivity~\citep{P_rez_Salinas_2020, schuld2021effect}.

Second, can the conditioning mechanism be learned end-to-end rather than manually specified? Parametrising and optimising the conditioning jointly with the generative model could lead to more efficient or task-specific encodings. This is akin to learned embeddings in classical conditional generative models such as classifier-free guidance~\citep{ho2022classifierfreediffusionguidance} or self-supervised learning~\citep{gui2024surveyselfsupervisedlearningalgorithms, Jaderberg_2022}.

As noted in the main text, employing local cost functions is crucial for enabling the efficient training of larger parametrised quantum circuits~\citep{mcclean2018barren, cerezo2021cost}. However, validating the outputs of these larger models classically remains a significant computational bottleneck for standard state-vector simulators. Consequently, a promising direction for future work is to simulate our architecture well beyond the current $n \leq 4$ qubit demonstrations. This could be achieved by leveraging classical simulation frameworks that avoid full Hilbert space representations, such as Matrix Product States (MPS)~\citep{Or_s_2014}, or the recently developed Pauli propagation and Heisenberg-picture tracking methods~\citep{Begu_i__2025, rudolph2025paulipropagationcomputationalframework}.

\begin{acknowledgments}
DQ acknowledges support from the UK EPSRC through Grant No. EP/W52444X/1 and is grateful for use of the computing resources from the Northern Ireland High Performance Computing (NI-HPC) service funded by UK EPSRC Grant No. EP/T022175. G.D.C. acknowledges financial support from the Spanish Agencia Estatal de Investigación project PID2024-162153NB-I00 and UKRI-EPSRC project UKRI3314‌. The authors thankfully acknowledge the computer resources at MareNostrum and the technical support provided by Barcelona Supercomputing Centre (FI-2025-2-0017).‌
\end{acknowledgments}

\section*{Data Availability}
The data that support the findings of this study will be openly available following an embargo at the following URL/DOI: \href{https://github.com/DanielQuinn53/CQDD}{https://github.com/DanielQuinn53/CQDD}~\citep{quinn2025cqdd}. Data will be available from 27 February 2026.

\appendix
\section{Comparison of Conditioning Methods}
\label{app:methods}

Table~\ref{tab:conditioning} presents the values of the final testing loss obtained using various conditioning methods across different multi-class generation tasks. For the 2-class problem, ancilla qubits are initialised to the computational basis states $\ket{00}$ and $\ket{11}$, with conditioning angles set to $0$ and $\pi$, respectively. In the 4-class case, the ancilla system is initialised to all four computational basis states, with conditioning angles of $0$, $\pi/2$, $\pi$, and $3\pi/2$. The problem denoted `equator rings' is the task as described in Section~\ref{subsec:problem-scale} given by Eq.~\eqref{eq:alpha}. For the entanglement task, we use the problem as discussed in Section~\ref{subsec:entnalge} for $n=2$ qubits.
\begin{table}
    \centering
    \setlength{\tabcolsep}{10pt}
    \begin{tabular}{cccc}
        \toprule
        \textbf{Problem} & \textbf{Anc.} & \textbf{RX} & \textbf{RY} \\
        \midrule
        2 Equator Rings & 11.1\% & \textbf{6.37}\% & 9.14\% \\
        4 Equator Rings & 4.33\% & \textbf{3.98}\% & 5.27\% \\
        Entanglement & 2.78\% & 4.12\% & \textbf{2.72}\% \\
        \bottomrule
    \end{tabular}
    \caption{\textbf{Comparison of Conditioning Methods.} Values of the final test loss $\mathcal{L}(0)$ for different conditioning approaches: ancilla-based initialisation (Anc.) and label encoding via rotation gates ($RX$ and $RY$). All experiments for each method use identical model hyperparameters with $n_a=2$ and consistent random seeds for data generation and training. Bold values indicate the lowest test loss for each problem.}
    \label{tab:conditioning}
\end{table} 

We omit the results using the $RZ$ rotation gate, as it performed significantly worse in all cases. This underperformance arises because $RZ(\mu)\ket{0} = e^{-i\mu/2}\ket{0}$ introduces only a global phase, which has no significant effects on the state. As a result, the conditioning label fails to encode meaningful class information, and the model is unable to distinguish between different input classes. Consequently, the model converges to a distribution that represents an average over all classes. For example, in the two-class equator ring scenario, where each class corresponds to a ring offset from the equator, the $RZ$ conditioning leads our CQDD model to generate a distribution centred around the equator, which was not explicitly part of the training data.  This behaviour is reminiscent of the \emph{mode collapse} in classical generative models, where the generator maps multiple inputs to the same output, leading to poor generalisation and low sample diversity~\citep{modecollaspe}. 

\section{Entanglement Rings and 3 Qubit Results}
\label{app:ent}

Visualising the full entanglement structure of two- or three-qubit quantum states is inherently non-trivial due to the high dimensionality of the Hilbert space. We employ a projected Bloch sphere representation focused on specific subspaces. For the $S_\Phi$ class from Eq.~\eqref{eq:phi}, we project the state onto the subspace spanned by $\{\ket{00}, \ket{11}\}$. This projection is isomorphic to a single-qubit state of the form $(\ket{0} + e^{i\phi}\ket{1})/\sqrt{2}$, and hence, we expect the Bloch sphere to exhibit a ring-like structure in the $XY$-plane.

We extend our approach to $n=3$ qubits and define two distinct target distributions. One class is $S_{\rm GHZ}$ with states $\ket{\mathrm{GHZ}} = \ket{000} + e^{i\phi}\ket{111}$ up to some normalisation, and the other class is $S_{\rm W}$ that includes states $\ket{\mathrm{W}} = \ket{001} + \ket{010} + e^{i\phi}\ket{100}$, again up to the appropriate normalisation. We train a model with $T=30$, $L=12$ and $N=250$.

As before, we employ a subspace percentage metric to evaluate the generation quality. Take a general 3 qubit state $\ket{\psi}=\sum_{i=1}^{8} c_i\ket{i}$, then the percentage of the state overlapping with the subspace of GHZ states is $|c_0|^2+|c_7|^2$, while for the W states is $|c_1|^2+|c_2|^2+|c_4|^2$. The model successfully captures both GHZ and W classes, with roughly 99\% of the generated samples lying in the correct subspace. However, the visual structure of the generated data differs significantly from the two-qubit case.

\begin{figure}[htbp]
    \centering
    \includegraphics[width=0.8\linewidth]{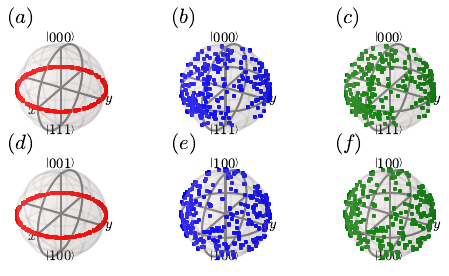}
    \caption{\textbf{Projected Bloch Spheres.} (a-c) Projection of the target, training, and testing GHZ states onto the subspace spanned by $\{\ket{000},\ket{111}\}$. (d-f) Projection of the target, training, and testing W states onto the subspace spanned by $\{\ket{001},\ket{100}\}$. All the plotted states have been renormalised.}
    \label{fig:ring3q}
\end{figure}

In Fig.~\ref{fig:ring3q}(a)-(c), we project the $S_{\rm GHZ}$ onto the subspace spanned by $\{\ket{000},\ket{111}\}$, which is isomorphic to a single qubit. Although the projection is faithful and 99\% of the state lies in the correct subspace, the training and testing distributions do not yield a clean ring. Similarly, the W-class states, which lie in the rank-3 subspace $\{\ket{001},\ket{010},\ket{100}\}$, do not yield a clean geometric structure upon projection, as shown in Fig.~\ref{fig:ring3q}(d)-(f). 

\section{Product and Entangled States}
\label{app:rank}

Contrary to naive expectations, learning product states is empirically more difficult than learning entangled states. We conjecture that this originates from the structure and effective rank of the states involved. Below, for simplicity, we explore the case of generating a single class of states.

Consider first the two-qubit case. A typical product state with a random local phase \(\phi\) forms the set
\begin{equation}
S_{\rm{PS}} = \left\{ (\ket{0}+e^{i\phi}\ket{1})^{\otimes 2} \quad \text{s.t.} \quad\phi \sim \mathcal{U}(0, 2\pi) \right\},
\end{equation}
whereas a set of entangled states with the same phase dependence can be written as
\begin{equation}
S_{\rm{Ent}} = \left\{ \frac{\ket{00}+e^{i\phi}\ket{11}}{\sqrt{2}} \quad \text{s.t.} \quad\phi \sim \mathcal{U}(0, 2\pi) \right\}.
\end{equation}
Expanding the product state, we see $(\ket{0}+e^{i\phi}\ket{1})^{\otimes 2}=\ket{00}+e^{i\phi}(\ket{01}+\ket{10})+e^{2i\phi}\ket{11}$ up to some normalisation. The product state features phase contributions of both \(\phi\) and \(2\phi\), while the entangled state depends only on a single phase. In this sense, the entangled state resides in a lower-dimensional manifold and is effectively governed by a single-qubit parameter. By contrast, the product state exhibits full rank over the Hilbert space and introduces multiple dependent phase components, thus increasing its representational complexity.

This effect becomes even more pronounced for three qubits. The corresponding set of product states is
\begin{equation}
S_{\rm{PS}} = \left\{ (\ket{0}+e^{i\phi}\ket{1})^{\otimes 3} \quad \text{s.t.} \quad\phi \sim \mathcal{U}(0, 2\pi) \right\},
\end{equation}
which expands as $\ket{000} + e^{i\phi}(\ket{001} + \ket{010} + \ket{100}) + e^{2i\phi}(\ket{011} + \ket{110} + \ket{101}) + e^{3i\phi} \ket{111}$ for a single term, up to some normalisation. The entangled state forms a set of GHZ states, 
\begin{equation}
S_{\rm{Ent}} = \left\{ \frac{\ket{000}+e^{i\phi}\ket{111}}{\sqrt{2}} \quad \text{s.t.} \quad\phi \sim \mathcal{U}(0, 2\pi) \right\}.
\end{equation}
Again, the product state distributes the phase across many amplitudes, while the entangled state retains a simpler structure. Empirical evidence supports this structural interpretation. For two qubits with \(n_a = 2\) ancilla qubits, $T=20$ diffusion steps, \(L = 10\) circuit depth and $N=250$ training states, the final test loss for the product state is approximately 7.44\%, compared to the 1.52\% for the entangled state (roughly 5 times larger). For three qubits under the same model settings, the loss for the product state is 22\%, compared to 3.74\% for the entangled state (roughly 6 times larger). These results suggest that higher-rank states with a more distributed phase structure are indeed harder to learn for the CQDD model.

\section{Quantum Direct Transport and Other Generative Models}
\label{app:direct}

Whilst the main text benchmarks the CQDD framework against an unconditioned QDD baseline, this appendix extends the analysis to compare to a conditioned quantum direct transport (CQDT) model. We then discuss other classes of quantum and hybrid generative models, although they are not implemented here. 

\subsection{Quantum Direct Transport}

The CQDT model may be regarded as a generalisation of the quantum circuit Born machine~\citep{Benedetti_2019, Liu_2018, Coyle_2020, herrerogonzalez2025bornultimatumconditionsclassical}. It employs the same dataset, conditioning mechanism, and loss function as CQDD, but replaces the multi-step diffusion process with a single transformation of depth $L_{\rm DT}$, mapping an initial ensemble of random quantum states directly to the target distribution.

To ensure a controlled comparison, both the variational parameter count and the total optimisation budget are matched across models. The CQDD implementation employs $T = 20$ diffusion steps, each parametrised by a quantum circuit of depth $L = 5$. Accordingly, the CQDT model is realised as a single parametrised quantum circuit of depth $L_{\rm DT} = 100$. In addition, the CQDT model is trained for $100{,}000$ epochs, aligning with the total number of optimisation updates of the CQDD model (20 steps of 5000 epochs).

Both architectures are evaluated on the ring dataset, targeting the generation of the state classes $S_X$ and $S_Y$ defined in Eqs.~\eqref{eq:sx} and \eqref{eq:sy}, using $n_a = 2$ ancilla qubits and $N = 100$ samples. Empirically, CQDD significantly outperforms its direct transport counterpart, achieving a testing error of $27.7\%$, compared to $95.5\%$ for CQDT. This performance gap of $3\times$ highlights a key advantage of diffusion-based approaches: by decomposing a complex state transformation into a sequence of intermediate, more tractable mappings, CQDD mitigates the optimisation challenges associated with training deep, single-step quantum circuits.

\subsection{Other Generative Models}

Beyond circuit Born machines and direct transport models, an important class of quantum generative frameworks is given by Quantum Generative Adversarial Networks (QGANs). This class of model has a adversarial training methodology, involving competing generator and discriminator circuits~\citep{Lloyd_2018, Hu_2019, PhysRevA.98.012324, PhysRevApplied.16.024051}. Whilst QGANs provide a conceptually distinct approach to distribution learning, a comparison with diffusion-based quantum models presents non-trivial challenges. In particular, differences in training dynamics, objective functions, and stability properties complicate direct benchmarking under matched resource constraints. For these reasons, we defer a detailed empirical evaluation of QGAN-based approaches to future work.

Finally, other quantum generative paradigms, most notably Quantum Boltzmann Machines (QBMs)~\citep{Amin_2018, crawford2019reinforcementlearningusingquantum, Zoufal_2021}, utilise energy-based objectives to align the thermal states of a given Hamiltonian with the target data distribution. These models require complex training procedures and are often constrained by the significant computational overhead associated with Gibbs state preparation. Given their reliance on energy-based sampling, QBMs are not directly comparable and are therefore omitted from the comparative analysis.

\section{Gradients and Complexity}
\label{app:grad}

In this appendix, we outline additional technical details regarding the computational complexity and gradient behaviour of our CQDD model.

\subsection{Runtime Complexity}

A defining feature of the CQDD model is the use of a sequential optimisation strategy, wherein an independent set of parameters is trained for each diffusion step $t$. This results in a classical training complexity that scales linearly with the number of diffusion steps, $\mathcal{O}(T)$. While this procedure is not trivially parallelisable, it is essential for maintaining trainability.

Training a single global parametrised circuit $\tilde{U}(\boldsymbol{\theta})$ would require sufficient expressibility to capture the entire diffusion trajectory. In practice, this would necessitate deep circuits, which are known to severely hinder optimisation (see Appendix~\ref{app:direct}). In contrast, the sequential approach effectively decomposes the learning problem into a collection of simpler tasks, each associated with a shallow circuit. Also, the computational overhead associated with this strategy is only confined to the training phase. During inference, state generation requires the execution of the learned reverse circuits, with the total quantum gate depth scaling as $\mathcal{O}(T L)$.

When using exact statevector simulation, the runtime complexity of the training procedure per epoch scales as
\begin{equation}
    \mathcal{O}\left(T \, L \, |\mathcal{C}| \, 2^{n + n_a} \, N^2 \right),
\end{equation}
where $T$ is the number of diffusion steps, $L$ is the circuit depth, $|\mathcal{C}|$ is the number of classes, $n + n_a$ is the total number of qubits (system plus ancilla), and $N$ is the dataset size. The exponential dependence on the number of qubits arises from statevector simulation, while the quadratic dependence on $N$ arises from the construction of the pairwise cost matrix required to compute the Wasserstein distance between the generated and target distributions.

\subsection{Gradient Stability}

In PQCs, increasing circuit depth $L$ or system size $n$ can lead to barren plateaus, characterised by an exponential suppression of gradients in the number of qubits~\citep{mcclean2018barren, cerezo2021cost}. This behaviour is typically associated with highly expressive circuits that approximate unitary $2$-designs, resulting in concentration effects and vanishing gradient signals~\citep{mele2024introduction}. To evaluate whether the proposed sequential optimisation strategy preserves a trainable landscape throughout the reverse process, we analyse gradient behaviour in the CQDD model.

At each diffusion step $t$, we compute the squared $L_2$ norm of the gradient of the loss function with respect to the trainable parameters $\boldsymbol{\theta}_t \in \mathbb{R}^d$. Concretely, we compute
\begin{equation}
    \|\nabla_{\boldsymbol{\theta}_t} \mathcal{L}(t)\|_2^2
    =\sum_{i=1}^d \left( \frac{\partial \mathcal{L}(t)}{\partial (\boldsymbol{\theta}_t)_i} \right)^2.
    \label{eq:grad_norm}
\end{equation}
This quantity provides a practical measure of the overall gradient signal during optimisation.

\begin{figure}[htbp]
    \centering
    \includegraphics[width=\linewidth]{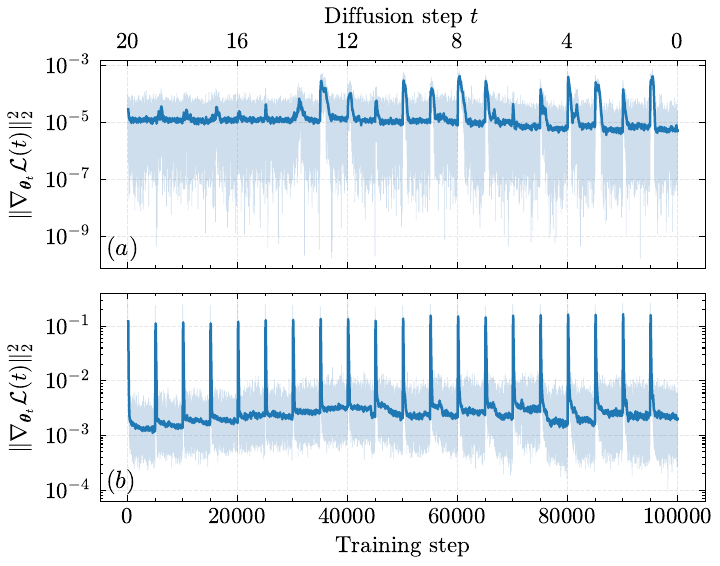}
    \caption{\textbf{Gradient Norm.} The squared $L_2$ norm of the gradients with respect to the circuit parameters during optimisation of the denoising process, for (a) $L=1$ and (b) $L=10$ circuit layers. The raw gradient values are shown as faint traces, while the solid blue curves correspond to a moving average with window size $100$, highlighting the underlying trend. The gradients remain non-vanishing and well-behaved throughout training, with no indication of decay or instability.}
    \label{fig:gradient_norm}
\end{figure}

Fig.~\ref{fig:gradient_norm} shows the evolution of the squared $L_2$ norm over training, where the task corresponds to generating the state classes $S_X$ and $S_Y$ defined in Eqs.~\eqref{eq:sx} and \eqref{eq:sy}, using $T=20$ diffusion steps, $n_a=2$ ancilla qubits, and $N=500$ samples. The raw gradient norms are shown as faint traces, while the solid blue curves correspond to a moving average with a window size of $100$. In Fig.~\ref{fig:gradient_norm}(a), we consider shallow circuits with $L=1$, while Fig.~\ref{fig:gradient_norm}(b) corresponds to deeper circuits with $L=10$. The periodic peaks observed are a direct consequence of the sequential training procedure, as each new diffusion step $t$ is initialised independently.

Notably, we observe a distinct shift in scale between the two depths. For $L=1$, the squared gradient norm drops to a range spanning approximately $10^{-9}$ to $10^{-3}$, indicating a shallow circuit rapidly converging to a rigid local minimum. Conversely, for $L=10$, the norm lies in a higher, narrower range of approximately $10^{-4}$ to $10^{-1}$. This increase is expected, as the deeper circuit encompasses a larger number of parameters $d$ in the sum of Eq.~\eqref{eq:grad_norm} and navigates a more expressive loss landscape, thereby maintaining a stronger gradient signal. Crucially, despite the increased depth, the $L=10$ model does not exhibit the rapid exponential suppression characteristic of barren plateau regimes. We therefore conclude that, within the explored settings, the sequential approach prevents gradient collapse and supports stable optimisation, albeit at the cost of increased classical training overhead.

\section{Hardware Deployment and Noise Considerations}
\label{app:noise}

This appendix details the dynamic circuit requirements, where dynamic circuits refer to the use of mid-circuit measurement and reset operations. We also analyse the model's sensitivity to both coherent and incoherent noise.

\subsection{Dynamic Circuits and Incoherent Noise}
To have a constant ancilla overhead across $T$ sequential diffusion steps, the hardware must support dynamic circuits, i.e. the quantum processor must implement mid-circuit measurements and high-fidelity resets to repeatedly prepare the ancilla subsystem with the conditioning parameter $\mu$ at each step. Such hardware demands make our model naturally suited for platforms like superconducting circuits or trapped-ion systems that are capable of intermediate measurements without destroying the coherence of data qubits~\citep{C_rcoles_2021, PhysRevX.13.041057, yu2025insitumidcircuitqubitmeasurement}. Conversely, if this hardware capability is not met, the sequential diffusion process would require continuously swapping in new, explicitly prepared ancilla qubits at each diffusion step, resulting in an overhead of $T\times n_a$ ancilla qubits.

In realistic implementations, the PQC is subject to cumulative noise processes, including depolarising errors from successive two-qubit gates, as well as amplitude damping and phase damping. These effects progressively degrade the trained channels $\tilde{U}(\boldsymbol{\theta}_t)$, resulting in the generation of mixed states that deviate from the target distribution. To ensure that the generated states retain sufficient fidelity, the incorporation of quantum error mitigation techniques is essential. In particular, methods such as Zero Noise Extrapolation and Probabilistic Error Cancellation are required to counteract the accumulation of depolarising errors arising from the $\mathcal{O}(TL)$ circuit depth~\citep{cai2023quantum}. In addition, measurement and reset errors affecting the ancilla register must be mitigated using readout correction protocols, such as Twirled Readout Error Extinction~\citep{van2022model} or Matrix-free Measurement Mitigation~\citep{Nation_2021}.

\subsection{Robustness to Coherent Noise}

To assess the model's resilience to hardware-level imperfections, we numerically investigate its sensitivity to coherent static noise. In physical implementations, such noise typically arises from miscalibrations in control fields (e.g., microwave pulse amplitudes, phases, or durations), leading to over- or under-rotations in variational circuits~\citep{ball2021software}.

We model these effects as static, gate-independent perturbations to the optimised parameters $\boldsymbol{\theta}_t$. Concretely, we apply the mapping
\[
\theta_{t,i} \;\mapsto\; \theta_{t,i} + \epsilon_i,
\]
where each $\epsilon_i$ is independently drawn from a zero-mean Gaussian distribution, $\epsilon_i \sim \mathcal{N}(0,\sigma)$. This corresponds to a static coherent error model, capturing a calibration drift that remains fixed over the course of the circuit execution.

\begin{figure}[htbp]
    \centering
    \includegraphics[width=0.8\linewidth]{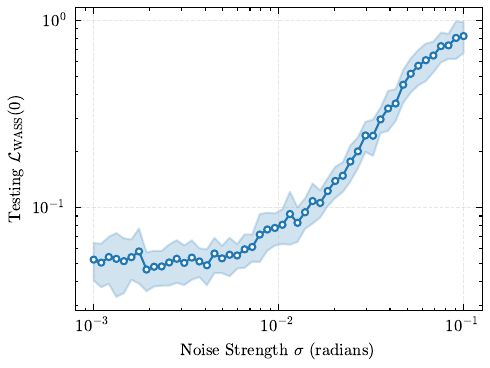}
    \caption{
    \textbf{Robustness to Coherent Noise.}
    Test loss as a function of the coherent noise strength $\sigma$, where static Gaussian perturbations $\epsilon_i \sim \mathcal{N}(0,\sigma)$ are applied to all circuit parameters. Results are shown for 10 random realisations of the perturbations, with the markers denoting the mean and the shaded region one standard deviation.}
    \label{fig:coherent_noise}
\end{figure}

Our numerical results, depicted in Fig.~\ref{fig:coherent_noise}, indicate that the CQDD loss landscape exhibits broad, flat minima, and consequently the model demonstrates robustness to small coherent perturbations. In the main text, and consistently across all simulations, trained models do not exceed a $10\%$ test loss and hence we adopt this as our practical tolerance threshold. From Fig.~\ref{fig:coherent_noise}, we observe that this criterion is satisfied for coherent noise strengths up to $\sigma = 10^{-2}$ radians. Beyond this threshold, performance degrades rapidly, and thus this transition marks an effective tolerance scale for coherent control errors. The reported results correspond to the generation of the state classes $S_X$ and $S_Y$ defined in Eqs.~\eqref{eq:sx} and \eqref{eq:sy}, using $T=20$ diffusion steps, $n_a=2$ ancilla qubits, circuit depth $L=12$, and $N=500$ samples.

\section{Architectural and Simulation Details}
\label{app:details}

Table~\ref{tab:details} shows the model hyperparameters used for each of the problems discussed in the main text. 

The simulation of the CQDD model is carried out using the Python library \texttt{TensorCircuit}~\citep{Zhang_2023}, while Bloch sphere visualisations are generated with the assistance of \texttt{QuTiP}~\citep{Johansson_2012}. Computation of the Wasserstein distance is performed using the Python Optimal Transport library \texttt{POT}~\citep{JMLR:v22:20-451}.

\begin{table*}[t]
    \centering
    \setlength{\tabcolsep}{8pt} 
    \begin{tabular}{lccccccccr}
        \toprule
        \textbf{Problem} & $|\mathcal{C}|$ &\textbf{$n$} & \textbf{$n_a$} & \textbf{$L$} & \textbf{$T$} & \textbf{$N$} & \textbf{$\delta_t$} & \textbf{Distance} & \textbf{Performance} \\
        \midrule
        Planar Rings (\ref{subsec:topological})      & 3 & 1   & 2   & 15  & 20  & 1000 & $0.005t^2$  & WASS & 4.3\% \\
        Polar Points (\ref{subsec:cardinal})     & 6 & 1   & 2   & 12  & 20  & 500  & $0.15t$ & MMD & 1.5\% \\
        Entanglement Structure (\ref{subsec:entnalge}) & 2 & 2 & 2 & 12 & 20 & 125 & $0.01t^2$ & WASS & 2\% \\
        Many Body Phases (\ref{subsec:manybody})     & 2 & 4   & 2 & 12 & 30 & 100 & 0.1 to 2 & MMD & 3.7\% \\
        \bottomrule
    \end{tabular}
    \caption{\textbf{Model Hyperparameters and Performance.} Summary of the CQDD model hyperparameters and its performance across various generative learning tasks. $|\mathcal{C}|$ indicates the number of different classes the model wishes to generate from. The number of data qubits is $n$, and $n_a$ denotes the number of ancilla qubits. $L$ is the depth of each individual denoising circuit $\tilde{U}$, and $T$ is the number of diffusion steps. $N$ represents the number of samples used during training and testing. $\delta_t$ refers to the noise schedule. The notation $x$ to $y$ implies a linearly spaced set of length $T$ starting for $x$ and ending at $y$. Distance denotes which metric we use, either maximum mean discrepancy (MMD) or Wasserstein (WASS). Performance is reported as the final test value of the loss function $\mathcal{L}(0)$.}
    \label{tab:details}
\end{table*}

\newpage

\bibliographystyle{apsrev4-2}
\bibliography{Refs-CQDD}

\end{document}